# Exploiting a Supervised Index for High-accuracy Parameter Estimation in Low SNR

Kaijie Xu

*Abstract*—Performance of parameter estimation is one of the most important issues in array signal processing. The root mean square error, probability of success, resolution probabilities, and computational complexity are frequently used indexes for evaluating the performance of the parameter estimation methods. However, a common characteristic of these indexes is that they are unsupervised indexes, and are passively used for evaluating the estimation results. In other words, these indexes cannot participate in the design of estimation methods. It seems that exploiting a validity supervised index for the parameter estimation that can guide the design of the methods will be an interesting and meaningful work. In this study, we exploit an index to build a supervised learning model of the parameter estimation. With the developed model we refine the signal subspace so as to enhance the performance of the parameter estimation method. The main characteristic of the proposed model is a circularly applied feedback of the estimated parameter for refining the estimated subspace. It is a closed loop and supervised method not reported before. This research opens a specific way for improving the performance of the parameter estimation by a supervised index. However, the proposed method is still unsatisfying in some scopes of signal-to-noise ratio (SNR). We believe that exploiting a validity index for the parameter estimation in array signal processing is still a general and interesting problem.

*Index Terms*—Array signal processing, Direction Of Arrival (DOA), Root Mean Square Error (RMSE), Signal subspace.

## I. Introduction

Parameter estimation in array signal processing [1] is concerned with the detection and estimation of signals and their parameters (includes direction of arrival, number of signals, frequency, etc.) from data collected by spatially distributed sensors or antennas [2]. It plays an important role in numerous applications [3], such as radar, sonar, wireless communication, radio astronomy, tomography and seismic exploration, etc. By exploiting the received data of the array and the properties of a steering matrix, a large number of approaches [4]–[6] have been developed to estimate the parameters of sources. The most representative methods are the Eigen-subspace methods [7].

The eigen-subspace methods, also known as high-resolution subspace algorithms, are commonly used in parameter estimation due to their high efficiency, high resolution, and high accuracy. This type of methods is sensitive to the precision of signal and noise subspaces, and the parameter estimation performance may significantly decrease when the signal and noise subspaces cannot be accurately estimated. There are several reasons behind this performance degradation. One of the most important reason is that when the signal-to-noise ratio (SNR) [8] is lower than 0 dB, the power of signal is equal or less than noise, signal and noise subspaces become difficult to separate. Hence an incorrect signal subspace is incurred.

In some scenarios, such as target identification, electronic countermeasures, stealth, remote, low altitude targets, the SNR is often much lower than 0 dB (say, lower than -15 dB), How to improve the performance of parameter estimation in very low SNR is a significant problem in most practically interesting situations.

However, improving the accuracy of signal and noise subspaces in low SNR is very difficult, i.e., the lower the value, the more difficult the task. It is well known that most of the parameter estimation algorithms would likely fail in low SNR because the signal-and-noise subspaces cannot be obtained accurately in this case.

In this work we propose a validity supervised index for the parameter estimation. Based on the reconstruction of the signal subspace [9], in this paper we exploit a validity supervised index for high resolution parameter estimation in low SNR. Unlike the traditional unsupervised indexes such as Root Mean Square Error (RMSE) [9], probability of success [10], [11], resolution ability [12], and computation efficiency [13], our proposed index can iteratively feedback the results of estimation, and cyclically optimize the signal subspace, and eventually it can enhance the performance of parameter estimation. Without loss of generality, we use the estimation of the Direction Of Arrival (DOA), which is an important parameter in array signal processing, for all the discussions and illustrations.

During the exploiting process, we establish a fuzzy similarity matrix for the eigenvalues of the correlation matrix of the array received data, and then build up a transformation matrix between the similarity matrix and the eigenspace of the correlation matrix. A non-linear transformation function is creatively introduced to adjust the similarity matrix. Accordingly, the estimated eigenspace can be modified with the transformation matrix and a new similarity matrix. Then a spectrum of DOAs is received with the modified signal subspaces. Subsequently, we define a reconstruction error of

K. Xu is with the School of Electronic Engineering, Xidian University, Xi'an 710071, China (e-mail: kjxu@xidian.edu.cn).

the signal subspace with which we build up a supervised learning model of the DOA estimation. Finally, the signal subspace is refined and therefore the DOA is optimized.

In the experimental section, we present several simulations to verify the proposed scheme. Simulations show that the proposed model provides excellent performance in low SNR (say, lower than -15 dB). To the best of our knowledge, the proposed scheme has not been considered in the previous studies.

This paper is structured as follows. The model of the signal is formulated in Section II. A series of evaluation indexes for the parameter estimation methods is discussed in Section III. A supervised learning model of DOA estimation with a new index is presented in detail in Section IV. Section VI includes experimental setup and analysis of completed results. Section V covers some conclusions.

## II. ARRAY SIGNAL MODEL

Consider $P$ narrowband far-field sources $s_p(t)$ $(p=1,2,\cdots,P)$ [14] impinging on a uniform linear antenna array (ULA) composed of $M$ antennas, where $t$ indexes the snapshot. Let $x(t)$ be the measured data of the $t$th measurement, and the ULA output at the $t$th snapshot is expressed in the following manner

$$\begin{aligned} \boldsymbol{x}(t) &= [x_1(t), x_2(t), \cdots, x_M(t)]^T \\ &= \boldsymbol{A}(\theta)\boldsymbol{s}(t) + \boldsymbol{n}(t) \\ \boldsymbol{s}(t) &= [s_1(t), s_2(t), \cdots, s_P(t)]^T \\ \boldsymbol{n}(t) &= [n_1(t), n_2(t), \cdots, n_M(t)]^T \end{aligned} \quad (1)$$

where $T$ stands for the transpose operation, $\boldsymbol{s}(t)$ and $\boldsymbol{n}(t)$ denote the vector of source signals and the noise vector, respectively. The latter is assumed to have a zero-mean and is independent of the observed signal. $\boldsymbol{A}(\theta)$ is the so-called array manifold matrix, which is described as follows

$$\boldsymbol{A}(\theta) = [\boldsymbol{a}(\theta_1), \boldsymbol{a}(\theta_2), \cdots, \boldsymbol{a}(\theta_p), \cdots \boldsymbol{a}(\theta_P)] \quad (2)$$

where $\theta_p$ is the $p$th DOA of the sources (relative to the array broadside), and

$$\boldsymbol{a}(\theta_p) = \exp\left[0, \cdots, j\frac{2\pi}{\lambda}d(m-1)\sin\theta_p, \cdots, j\frac{2\pi}{\lambda}(M-1)d\sin\theta_p\right]^T \quad (3)$$

is the steering vector of the $p$th source, $\lambda$ is the wavelength of the signals, $d$ is the spacing between adjacent antenna elements.

The eigenvalue decomposition of the array output correlation matrix

$$\boldsymbol{R}_x = E\left[\boldsymbol{x}(t)\boldsymbol{x}^H(t)\right] = \boldsymbol{A}\boldsymbol{R}_s(0)\boldsymbol{A}^H + \sigma^2\boldsymbol{I} \quad (4)$$

$\boldsymbol{R}_x$ can be written compactly in the following way:

$$\boldsymbol{R}_x = \boldsymbol{U}\boldsymbol{\Lambda}\boldsymbol{U}^H = \boldsymbol{U}_s\boldsymbol{\Lambda}_s\boldsymbol{U}_s^H + \boldsymbol{U}_n\boldsymbol{\Lambda}_n\boldsymbol{U}_n^H \quad (5)$$

where $H$ stands for the complex conjugate transpose, $\boldsymbol{R}_s = E[\boldsymbol{s}(t)\boldsymbol{s}^H(t)]$ is source correlation matrix, $\boldsymbol{U} = [\boldsymbol{u}_1, \boldsymbol{u}_2, \cdots, \boldsymbol{u}_P, \boldsymbol{u}_{P+1}, \cdots, \boldsymbol{u}_M]$ is the eigenvector of the correlation matrix, $\boldsymbol{\Lambda} = diag\{\xi_1, \xi_2, \cdots, \xi_P, \xi_{P+1}, \cdots, \xi_M\}$ is a diagonal matrix, $\boldsymbol{u}_m$ and $\xi_m$ $m=1,2,\cdots,M$ are the eigenvalue and corresponding eigenvector. The signal subspace $\boldsymbol{U}_s$ is spanned by the eigenvectors corresponding to $P$ largest eigenvalues, and noise space $\boldsymbol{U}_n$ is spanned by the eigenvectors corresponding to the other ($M$-$P$) eigenvalues. $\boldsymbol{\Lambda}_s$ and $\boldsymbol{\Lambda}_n$ are diagonal matrices whose diagonal elements are $\xi_1, \xi_2, \cdots, \xi_P$ and $\xi_{P+1}, \xi_{P+2}, \cdots, \xi_M$ respectively.

Note that the exact knowledge of $\boldsymbol{R}_x$ is unavailable, and thus in practical situations its estimation by sample averages is employed. This entails the following estimate

$$\hat{\boldsymbol{R}}_x = \frac{1}{N}\sum_{n=1}^{N}\boldsymbol{x}(t_n)\boldsymbol{x}^H(t_n) \quad (6)$$

where $N$ denotes the number of available snapshots. Then the estimated values of the corresponding signal and noise subspace are $\hat{\boldsymbol{U}}_s$ and $\hat{\boldsymbol{U}}_n$, and the DOA can be estimated based on the estimated signal and noise subspace.

Most of the existing high-resolution methods for the estimation of DOA are based on the numerical characteristics behind the entire array output correlation matrix [15].

## III. EVALUATION INDEXES

In this section, we recall a number of commonly used evaluation indexes in array signal processing. Without loss of generality, we use the DOA estimation as an example for all the illustrations.

### A. Root mean square error

The Root Mean Square Error (RMSE) characterizes the differences between the true and estimated DOAs, and as such, serves as a measure of the overall reliability of the method [9]. Typically, RMSE is defined as

$$\sqrt{\frac{1}{N_e}\sum_{n=1}^{N_e}\left\{\frac{1}{P}\sum_{p=1}^{P}\left[\hat{\theta}_p(n) - \theta_p\right]^2\right\}} \quad \text{(degree)} \quad (7)$$

where $N_e$ represents the number of independent trials, $\theta_p$ is the $p$th angle of the $P$ impinging sources, and $\hat{\theta}_p$ is the estimated value of $\hat{\theta}_p$ in the $n$th trial. The root mean square error (RMSE) criterion is the most frequently used index in signal processing.

### B. Probability of success

The probability of success (detection probability, success rate) [17], [21] characterizes the differences between the true and estimated angles, and as such, serves as a measure of the overall reliability of the algorithm. Usually, it can be expressed in the following manner

$$S_r = \frac{1}{N_e}\sum_{n=1}^{N_e}\frac{1}{P}\sum_{p=1}^{P}u\left(\delta - \left|\hat{\theta}_p(n) - \theta_p\right|\right) \quad (8)$$

where $u(t)$ is a unit step function, namely:

$$u(t) = \begin{cases} 1 & t \geq 0 \\ 0 & t < 0 \end{cases} \quad (9)$$

where $N_e$ is the number of independent trials, $\delta > 0$ is the success threshold, $\theta_p$ is the $p$th angle of the impinging sources, and $\hat{\theta}_p$ is the estimation of $\theta_p$ for the $n$th trial.

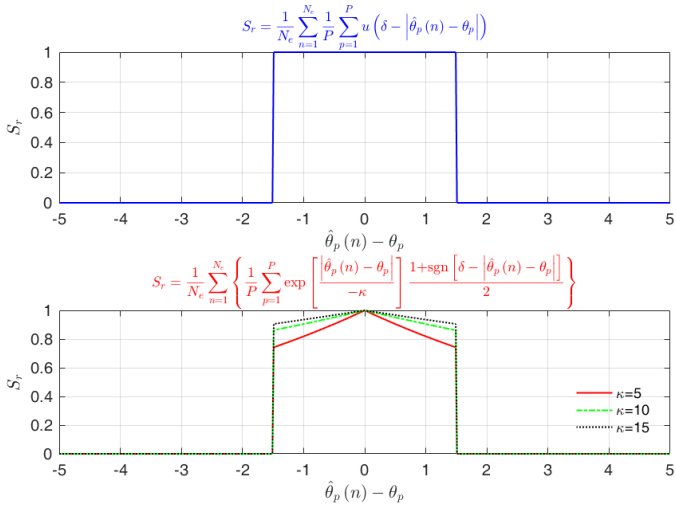

Fig. 1. Plots of curves of success rate.

Obviously $u(t)$ can be regarded as a two-valued logic function (Boolean) [22]–[23]. That is, successful detection indicates that the differences between the true and estimated angles are positioned within a certain margin $\delta$. In this sense, there is no difference between $|\hat{\theta}(n_1) - \theta| = \delta_1$ and $|\hat{\theta}(n_2) - \theta| = \delta_2$ ($\delta_1 < \delta_2 \leq \delta$) is taken into consideration. In fact, these results are different; apparently $\hat{\theta}(n_1)$ is closer to the true value. It is rational to reflect these differences present in these results, however the two-valued nature of the evaluation function $u(t)$ ignores such differences. Bearing this in mind, it could be beneficial to gain a closer insight into the quantification of the probability of success and refine the previous description. To accomplish this, we replace the unit step function $u(t)$ by its continuous counterpart to improve the traditional probability of success, which can be expressed as follows

$$S_r = \frac{1}{N_e}\sum_{n=1}^{N_e}\left\{\frac{1}{P}\sum_{p=1}^{P}\exp\left[\frac{|\hat{\theta}_p(n) - \theta_p|}{-\kappa}\right]\frac{1+\text{sgn}\left[\delta - |\hat{\theta}_p(n) - \theta_p|\right]}{2}\right\} \quad (10)$$

where $\kappa > 0$ is a coefficient of probability of success, which controls the rate of decline of the success rate function. This expression shows that, within the success rate threshold, the less the estimated value deviates from the true value, the greater the probability of success becomes. As this deviation increases, the probability of success rapidly declines. Once the estimated value deviates from the true value and exceeds some threshold, the estimate is identified as a failure. Obviously, the improved success rate is a "soft" index providing more flexibility than the traditional hard one. Fig. 1 visualizes and contrasts the behavior of these two indexes.

### C. Resolution probabilities (RP)

Two signals with DOA $\theta_1$ and $\theta_2$ are said to be resolved if the respective estimates $\hat{\theta}_1$ and $\hat{\theta}_2$ are such that both $|\hat{\theta}_1 - \theta_1|$ and $|\hat{\theta}_2 - \theta_2|$ are smaller than $|\theta_1 - \theta_2|/2$, as seen in [24].

### D. Computational complexity

The computational complexity of the algorithm can be expressed by the number of multiplication operations and addition operations (division is treated as multiplication, and subtraction is treated as addition) [25].

All of these indexes are commonly used to evaluate the quality of the estimated parameters of signals. However, they can only be used after the parameters are estimated, and they cannot optimize the estimated results (i.e., cannot participate in the design of the parameters estimation methods). In the following section we attempt to exploit an index which can optimize the estimation of parameters.

## IV. A SUPERVISED LEARNING MODEL OF DOA ESTIMATION WITH RECONSTRUCTION OF THE SIGNAL SUBSPACE

High-resolution subspace-based methods are popular parameter estimation algorithms in array signal processing applications due to their high efficiency, high resolution, and high accuracy [26]. The performance of most high-resolution methods mainly depends on the accuracy of the signal subspace. If the signal subspace is inaccurately estimated, high-resolution subspace-based algorithms would most likely fail. Thus, determining an accurate signal subspace is a meaningful work.

In this section, based on the array signal model, we illustrate a supervised learning model of DOA estimation and mine the information hidden in array data so as to refine the traditional signal subspace.

Assume that the resulting DOA is $\tilde{\theta}_p$ ($p = 1, 2, \cdots, P$). First, we can substitute $\tilde{\theta}_p$ ($p = 1, 2, \cdots, P$) to the array manifold matrix. Then the reconstruction manifold matrix $\tilde{A}$ is derived, with which we build up a signal subspace $\tilde{U}_s$ in the following form

$$\begin{aligned}\tilde{U}_s &= \tilde{A}(\tilde{A}^H\tilde{A})^{1/2}\\ \tilde{A} &= \left[a(\tilde{\theta}_1), a(\tilde{\theta}_2), \cdots, a(\tilde{\theta}_p), \cdots a(\tilde{\theta}_P)\right]\end{aligned} \quad (11)$$

Ideally, this signal subspace and the estimated signal subspace $\hat{U}_s$ based on the array output covariance matrix are equal. However, there always exists an error between them. We find [11] that this error of signal subspace reconstruction can be regarded as an index to measure the quality of the estimated DOA, i.e., the smaller the error is, the higher the accuracy of DOA becomes. By referring to the error of signal subspace, we define the signal subspace reconstruction error as

$$R_e = \left\|\tilde{U}_s\tilde{U}_s^H - \hat{U}_s\hat{U}_s^H\right\|_F \quad (12)$$

Based on the signal subspace reconstruction error, we first design a scheme to refine the signal subspace, reduce the reconstruction error, and increase the performance of the parameter estimation.

Define an $M$ row $M$ column square matrix $B$ as the fuzzy similarity matrix [27] of the eigenvalues of correlation matrix $\hat{R}_x$, where $b_{i,j} \in [0, 1]$ is the $i$th row and $j$th column element of $B$. We call $b_{i,j}$ as the fuzzy similarity degree between the $i$th and the $j$th eigenvalues calculated in the following form

$$b_{ij} = 1 - |\xi_i - \xi_j|\tau$$
$$\tau = \frac{1}{1 + \max(\xi_m) - \min(\xi_m)} \quad (13)$$
$$i, j, m = 1, 2, \cdots, M$$

where $\xi_i$ is the $i$th eigenvalue. In low SNR, $\xi_i \neq \xi_j (i \neq j)$, $\mathbf{B}$ is usually a full rank matrix; thus, there exists a matrix $\boldsymbol{\Omega}$ between the similarity matrix and the estimated eigen subspace $\hat{\mathbf{U}} = [\hat{\mathbf{U}}_s, \hat{\mathbf{U}}_n]$ that satisfies the following expression:

$$\hat{\mathbf{U}} = \mathbf{B}\boldsymbol{\Omega} \quad (14)$$

The matrix $\boldsymbol{\Omega}$ can be determined as follows:

$$\boldsymbol{\Omega} = (\mathbf{B}^T \mathbf{B})^{-1} \mathbf{B}^T \hat{\mathbf{U}} \quad (15)$$

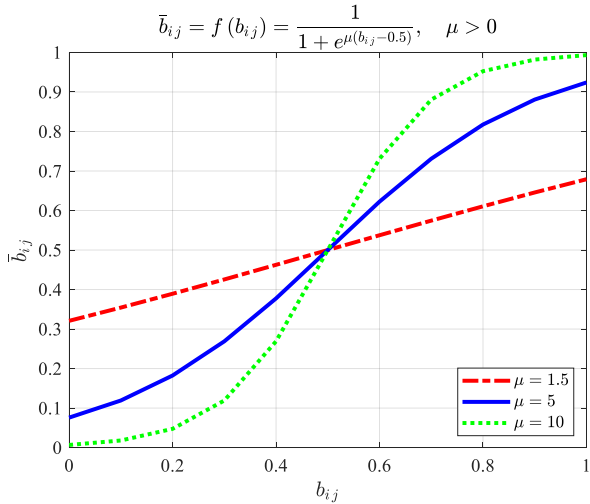

Fig. 2. Relationship between elements $b_{ij}$ and $\bar{b}_{ij}$ in the similarity matrices $\mathbf{B}$ and $\bar{\mathbf{B}}$.

It seems that Eq. (14) provides a direct adjustment mechanism for the eigen subspace $\hat{\mathbf{U}}$ and an indirect one for the signal and noise subspaces. Then we can consider to modify matrix $\mathbf{B}$ to adjust the signal subspace $\hat{\mathbf{U}}_s$. In this study, we construct such a modification function to adjust the elements in $\mathbf{B}$, which is described as

$$\bar{b}_{ij} = f(b_{ij}) = \frac{1}{1 + e^{\mu(b_{ij} - 0.5)}}, \quad \mu > 0 \quad (16)$$

where $\mu$ is a modification factor of the similarity degree. With the change of the modification factor $\mu$, a series of modified similarity matrices $\bar{\mathbf{B}}$ are obtained. The relationship between elements $b_{ij}$ and $\bar{b}_{ij}$ is plotted in Fig. 2. Then according to (14), a collection of modified subspaces $\hat{\mathbf{U}}$ is generated, with which several groups of DOAs can be determined. Through the use of the defined signal subspace reconstruction error $R_e$, we establish a composite cost function in the form:

$$J(\mu) = \left\| \tilde{\mathbf{U}}_s(\tilde{\theta}_p) \tilde{\mathbf{U}}_s^H(\tilde{\theta}_p) - [\bar{\mathbf{B}}(\mu)\boldsymbol{\Omega}]_P [\bar{\mathbf{B}}(\mu)\boldsymbol{\Omega}]_P^H \right\|_F$$
$$J(\mu) = \left\| \tilde{\mathbf{U}}_s[\tilde{\theta}_p(\mu)] \tilde{\mathbf{U}}_s^H[\tilde{\theta}_p(\mu)] - [\bar{\mathbf{B}}(\mu)\boldsymbol{\Omega}]_P [\bar{\mathbf{B}}(\mu)\boldsymbol{\Omega}]_P^H \right\|_F \quad (17)$$

where $[\bar{\mathbf{B}}(\mu)\boldsymbol{\Omega}]_P$ denotes the subspace spanned by the eigenvectors corresponding to $P$ largest eigenvalues. We have to determine the optimal (an acceptable) $\mu$ (by selecting a similarity matrix) to refine the subspace $\hat{\mathbf{U}}$ and optimize the estimated DOA so as to minimize the composite cost function. Obviously, the optimal modification factor can be solved by a searching method (To achieve the above requirements in this study we adopt the searching method.) Fig. 3 shows the principle of the proposed scheme.

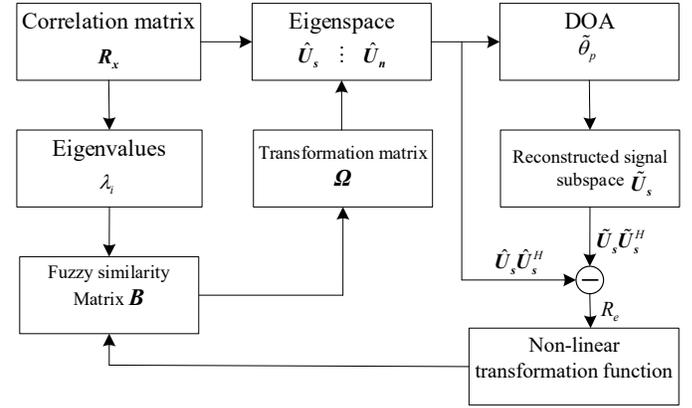

Fig. 3 An overall model: main functional processing phases.

## V. Experimental Studies

In this section, we present some simulations to test the performance of the proposed scheme. In the simulations, we mainly focus on the low SNR scenarios. An eight-element ULA with a relative inter-element spacing $d = \lambda/2$ is considered, and three narrowband source signals impinge on the ULA. We assume that the wave directions of the signals are 15°, 30°, and 45°, respectively. In the simulation the MUSIC algorithm is used to estimate the DOA, and the search step is set as 0.1.

**Simulation 1**

In the first simulation, we test the relationship between the RMSE and the proposed signal subspace reconstruction error index. The SNR is set as -18dB, and 50 snapshots are taken. The parameter $\mu$ in the modification function is set from 0 to 5 with 0.1 interval. A total of 200 Monte Carlo experiments are completed, and the means of the simulation results are plotted in Fig. 4.

It is apparent that the RMSE and the signal subspace reconstruction error basically have the same changing tendency, which demonstrates that, on the one hand, the new index can be used to evaluate the performance of the parameter estimation methods; on the other hand, the quality of the parameter estimation can be enhanced through modifying the eigenvector of the correlation matrix (eigenspace) with the supervision of the proposed new index.

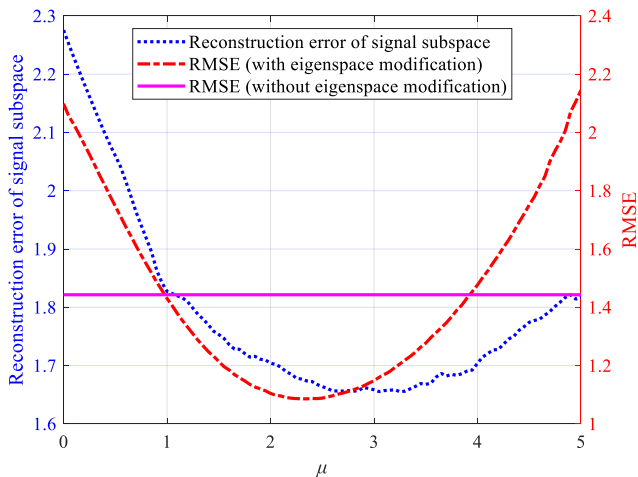

Fig. 4. The relationship between the RMSE and the signal subspace reconstruction error.

**Simulation 2**

In the second simulation, we test the RMSE performance of the proposed scheme and the MUSIC method versus SNR, where the snapshots are fixed at 50, and the SNR varies from -25 dB to -15 dB with 1 dB interval. The RMSE of the two algorithms versus SNR is depicted in Fig. 5. It is apparent that the proposed method outperforms the MUSIC method, and becomes insensitive to the changes of SNR. This simulation proves again that the proposed model is effective.

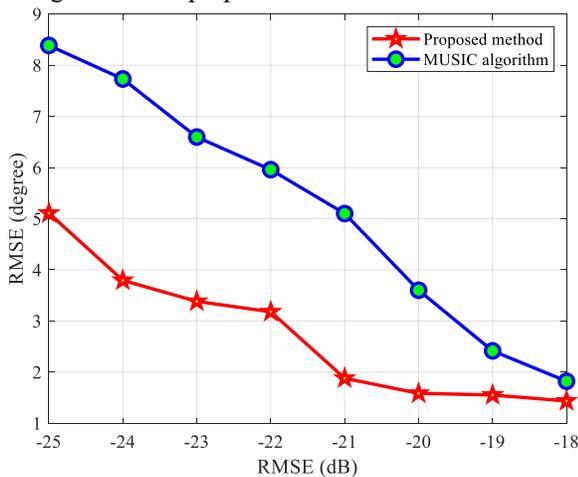

Fig. 5. Comparison of RMSE performance versus SNR.

## VI. CONCLUSIONS

In this research, we exploit a novel index for parameter estimation, with which we design an augmented model for DOA estimation. The research mainly focuses on improving the performance of DOA estimation in low SNR.

From the design perspective, the process revolves around the construction of a fuzzy similarity matrix and the optimization of the eigenspace of the correlation matrix. A novel index named signal subspace reconstruction error is introduced to supervise the optimization process. And the quality of the DOA is efficiently enhanced through optimizing the signal subspace. However, the new index is also somewhat limited as can be seen from the relationship between the RMSE and the signal subspace reconstruction error, i.e., their trends are not strictly consistent.

In short, this research opens a specific way for improving the performance of the parameter estimation and also poses a much general problem: How to exploit a validity supervised index for the parameter estimation in array signal processing? This question has also puzzled us for a long time. It is another reason that motivates us to propose this work in order to seek more insightful contributions from the researchers all over the world.

At the current stage, we have completed a theoretical analysis and offered a comprehensive suite of experiments. Some practical experiments (including hardware) would be an interesting avenue to explore in future studies.